\documentclass[final,3p,times,floatfix,amsmath,amssymb,preprintnumbers]{elsarticle}

\usepackage{graphicx}
\usepackage[brazil]{babel}
\usepackage{graphicx}
\usepackage{dcolumn}
\usepackage{bm}
\usepackage{subfigure}
\usepackage{listings}
\usepackage[all]{xy}
\usepackage{amsfonts}
\usepackage{amssymb}
\newcommand{\bef}{\begin{figure}}
\newcommand{\eef}{\end{figure}}
\newcommand{\bec}{\begin{center}}
\newcommand{\eec}{\end{center}}
\newcommand{\be}{\begin{equation}}
\newcommand{\ee}{\end{equation}}

\usepackage{amssymb}
\usepackage{multirow}

\usepackage{booktabs}
\usepackage{chngpage}
\usepackage{float}
\usepackage{setspace}
\RequirePackage{graphicx,subfigure}

\graphicspath{{figures/}}

\begin{document}

\begin{frontmatter}



\title{Smooth deformations and cosmic statefinders}

\author[A.J.S Capistrano1]{A.J.S Capistrano}
\address[A.J.S Capistrano1]{Federal University of Latin-American Integration, 85867-970, Foz do Igua\c{c}u-PR, Brazil}
\ead{abraao.capistrano@unila.edu.br}

\begin{abstract}
We study the possibility  that the  universe  is  subjected to  a  deformation,   besides its  expansion   described by Friedmann's  equations.  The concept of   smooth   deformation   of a  riemannian manifolds  associated  with  the  extrinsic  curvature  is applied the  standard  FLRW cosmology. Starting from the resulting modified Friedman's equation we study two possible solutions with six models for each one in low redshift. In other to constrain the models, we calculate deceleration, jerk and Hubble parameters and compare with different data as the latest BAO/CMB + SNIa constraints, SNLS SNIa, x-ray galaxy clusters and the gold sample (SNIa). As a result, we obtain a set of proper models compatible with the current observational data.
\end{abstract}


\end{frontmatter}

\section{Introduction}
The comprehension of the nature of dark matter and dark energy have been considered one of the greatest challenges in contemporary physics. In order to unravel these problems, the understanding of gravitational perturbation has been revealed to be an important issue. Recent observations of the CMBR power spectrum from the Planck mission \citep{planck} tell that the gravitational field perturbations amplify the higher acoustic modes due to the gravitational field of baryons. It is conjecture that such perturbations have to do with the influence of dark matter at cosmological scale that would induce a strong gravitational effect leading to a large scale structure formation. Similar effects of  gravitational perturbations are also verified at scales of galaxies and clusters, the so-called rotation curve problem.

In the same fashion, gravitational perturbations have influence on the proper explanation of dark energy that requires a more fundamental underlying theory. The  $\Lambda$CDM  model is interpreted  as  the   main  model for describing the acceleration expansion of the universe with the equivalence of quantum vacuum energy and cosmological constant. Besides its simplicity and consistency with the current observational data, it ignores the large  difference  between  the  very  small  observed  value  of the cosmological constant $\Lambda/8\pi G \approx  10^{-47} Gev^2/c^4$ and the  very  large averaged  value of quantum vacuum  energy  density  $<\rho_v> \approx  10^{75}  Gev^2/c^4$. Thus, the   absence of  a  feasible  solution makes  the  $\Lambda$CDM paradigm  an  improbable   explanation  of  the  accelerated  expansion of  the universe.  This  has motivated  the  emergence of   a  variety  of  alternative explanations, including   the possible  existence of new and  previously  unheard of  essences;  the  postulation of   specific   scalar   fields;  or  the  possible existence  of   non  observable    extra  dimensions in space.

In the recent years the extra dimensional  proposition seems to be an alternative route to solve the  hierarchy of the fundamental  interactions:   the   huge ratio of  the   Planck to   the electroweak energy  scale ($M_{Pl}/M_{EW} \sim 10^{16}$).   Since  Newton's  gravitational  constant $G$   depends on the dimension of  space, then   in  a    higher  dimensional  space   the    constant  $G$   must  change   to  another  value $G_*$,  such that  gravitating  masses   can be correctly   evaluated  by a (higher  dimensional) volume integration of   mass  densities. However,   the  existence  of  extra  dimensions must be compatible  with  the  experimentally proven  and  mathematically  consistent  four-dimensionality  of   space-times. This compatibility can be achieved  by assuming  that  the  space-times must  remain  four-dimensional,   but they are   subspaces  of   a higher-dimensional  space with  geometry defined  by  the  Einstein-Hilbert principle.   In this case, although  the  gravitational field   propagates  in  the  extra  dimensions, the  standard gauge  interactions  and  ordinary matter  remain confined  to  the four-dimensional space-time acting as    a  domain  wall  \citep{Rubakov}.

Several models  have  been   proposed  along  this  direction,  mostly  belonging  to  the  brane-world  paradigm  originally proposed in  \citep{ADD},  sometimes  using additional  conditions \citep{RS,RS1},  or  other  specific   embedding  assumptions  as for  example  in  e.g.~\citep{DGP,BW1,BW2,BW3,BW4,BW5,BW6,BW7,BW8,BW9}. In  spite  of such efforts we still do  not have a  complete  model  independent  solution for the  present  cosmological  problems \citep{Goenner}. The only theoretical relation between the extrinsic curvature and the material sources is the Israel-Lanczos boundary condition. However, it is well known that such a condition is of algebraic nature, so that it does not follow the evolution of the brane-world \citep{brandon}. Some similar models \cite{sepangi,sepangi1} have been developed with no need of particular junction conditions and/or with different junction conditions which lead to several approaches of brane-world models widely studied in literature \citep{maeda,anderson,sahni,Tsujikawa,gong}.

In a different approach, Nash's embedding theorem \citep{Nash} has been revealed to be a powerful tool on the understanding of both embedding process and evolution of a pseudo-riemannian/riemannian geometry (e.g, a braneworld spacetime) by using the concept of smooth deformations of the embedded geometries leading to formation of new ones. In five dimensions, a lesser known principle called Gutpa's theorem \citep{Gupta} is necessary to complement the dynamics of the extrinsic curvature originated from the Gauss-Codazzi equations. Based on the very foundations of geometry, the cosmology of \emph{smooth deformations} has been studied in a previous communication \citep{gde}. It was shown that the extrinsic curvature assumed a fundamental role of driving the propagation of gravitation along the extra dimensions of the bulk space and the most evident observable effect is the accelerated expansion of the universe.

In next section, we show the main results of the reference \citet{gde}. In the present paper, we explore two possible solutions for the Friedmann's equation called $\gamma^{(+)}$ and $\gamma^{(-)}$ in the low redshift range $0< z < 2.3$. In the section 3, we present the main purpose of this paper focusing on the study of the Hubble parameter and cosmic statefinders. The main cosmography parameters (deceleration and jerk parameters \citep{Visser1,Visser2}) to be studied in this work are applied to six models of each $\gamma$-solutions. For the resulting deceleration parameters, we compare with the latest BAO/CMB + SNIa constraints \citep{Giostri}. For the resulting jerks, we compare with SNLS SNIa \citep{astier}, x-ray galaxy clusters \citep{rapetti} and the gold sample (SNIa)\citep{riess}. In addition, in order to study the evolution of the Hubble parameter, we compare with the observational data extracted from \citep{cao} based on observations of red-enveloped galaxies \citep{stern} and BAO peaks \citep{gaztanaga} supplemented with the observational Hubble parameter data (OHD) with BAO in Ly$\alpha$ \citep{busca,zhai}. Finally, in the conclusion section, we present the final considerations.

\section{Einstein-Gupta's equations in the FLRW universe}
\subsection{Theoretical structure}
As shown in \citet{gde}, the Friedmann-Lema\^{\i}tre-Robertson-Walker(FLRW)
universe can be regarded as a brane-world in a five-dimensional bulk
with constant curvature. The constant curvature bulk is
characterized by the Riemann tensor
$$\mathcal{R}_{ABCD}=
K_{\ast}\left(\mathcal{G}_{AC}\mathcal{G}_{BD}-\mathcal{G}_{AD}\mathcal{G}_{BC}\right),
\;\;\;A.. D =  1 \cdots 5\;,$$ where $\mathcal{G}_{AB}$ denotes the
metric components of the bulk in arbitrary dimensions and $K_{\ast}=
6\Lambda$ is either zero(flat bulk) or it can be related to a
positive (deSitter) or negative (Anti-deSitter) a bulk cosmological
constant.

The cosmological observations \citep{riess} suggest that the
accelerated expansion scenario is compatible with the deSitter
$dS_5$ space-time. Moreover, the FLRW standard cosmological model is
completely embedded in a five-dimensional bulk. Thus, one can obtain the brane-world five-dimensional equations
\begin{equation}\label{1}
R_{\mu\nu}-\frac{1}{2}Rg_{\mu\nu}+\Lambda g_{\mu\nu}-Q_{\mu\nu}=
-8\pi G T_{\mu\nu}\;,
\end{equation}
\begin{equation}\label{2}
k_{\mu;\rho}^{\;\rho}-h_{,\mu} =0\;,
\end{equation}
which are the \emph{gravi-tensor} equation and \emph{gravi-vector} equation, respectively. The Greek indices vary from 1 to 4. The symbol $(;)$ represents the covariant derivative.

The  confinement  is set simply as $\alpha_* T^* =  8\pi G
 T_{\mu\nu}$, where $T_{\mu\nu}$ represents the energy momentum
tensor  of the confined  matter, and  $T^*_{\mu 5}=T^*_{55}=0$. Moreover, the quantity $Q_{\mu\nu}$ is given by
\begin{equation}\label{3}
  Q_{\mu\nu}=g^{\rho\sigma}k_{\mu\rho }k_{\nu\sigma
}- k_{\mu\nu }H -\frac{1}{2}\left(K^2-H^2\right)g_{\mu\nu}\;,
\end{equation}
where $k_{\mu\nu}$ denotes the extrinsic  curvature  and $H^2=
h.h\;,\;\;h=
 g^{\mu\nu}k_{\mu\nu}$, $K^{2}=k^{\mu\nu}k_{\mu\nu}$ . It is important to note that the quantity $Q_{\mu\nu}$ is a completely geometrical conserved quantity such that
\begin{equation}\label{4}
  Q_{\mu\nu;\nu}=0\;.
\end{equation}
In principle, the extrinsic curvature can be determined by Codazzi's
equation
\begin{equation}\label{5}
  k_{\mu[\nu;\rho]}=0\;,
\end{equation}
where brackets apply to the adjoining indices only. The details in obtaining these equations can be found in \citep{GDEI,gde,BW9,sepangi,sepangi1}.

In order to understand the influence of the extrinsic curvature on a FLRW
universe in a five-dimensional bulk, we take the line element in
$(r,\theta,\phi,t)$ coordinates such that
\begin{equation}\label{6}
ds^2=\;-dt^2+a^2\left[dr^2+f_k^2(r)\left(d\theta^2+\sin^2\theta
d\varphi^2\right)\right]\;,
\end{equation}
where $f(r)_k=\sin r$, r,$\sinh r$ that corresponds to spatial
curvature $\kappa$ = 1, 0, -1, respectively. As the cosmological observations
indicate an universe approximately flat, we adopt $k=0$ and
$f_k(r)=r$. The energy-momentum tensor $T_{\mu\nu}$ is the confined
source of a perfect fluid in co-moving coordinates given by
$$T_{\mu\nu}=(p+\rho)U_{\mu}U_{\nu}+p\;g_{\mu\nu},\;\;\;U_{\mu}=\delta_{\mu}^{4}\;.$$

After solving Codazzi's equations, one can obtain
\begin{eqnarray}
 &&k_{ij}=\frac{b}{a^2}g_{ij},\;\;i,j=1,2,3,
 \nonumber\\
&&k_{44}=\frac{-1}{\dot{a}}\frac{d}{dt}\frac{b}{a}\;,\nonumber\\
&&k_{11,\;\nu}=0\;;\;k_{11}=b(t)\;,\nonumber
\end{eqnarray}
where $b=b(t)=k_{11}$ is an arbitrary function of time \emph{t} and
$a=a(t)$ is the expansion parameter of the universe. Thus, one can
obtain
\begin{eqnarray}
\label{eq:BB}
 &&k_{ij}=\frac{b}{a^2}g_{ij},\;\;
 k_{44}=-\frac{b}{a^{2}}(\frac{B}{H}-1)g_{44},\;\;i,j=1..3,  \\
&&K^{2}=\frac{b^2}{a^4}\left( \frac{B^2}{H^2}-2\frac BH+4\right),
 \;\;\, h=\frac{b}{a^2}(\frac BH+2),\label{eq:hk}\\
&&Q_{ij}= \frac{b^{2}}{a^{4}}\left( 2\frac{B}{H}-1\right)
g_{ij},\;Q_{44} = -\frac{3b^{2}}{a^{4}}, i,j =1..3,
  \label{eq:Qab}\\
&&Q= -(K^2 -h^2) =\frac{6b^{2}}{a^{4}} \frac{B}{H}\;, \label{Q}
 \end{eqnarray}
where the dot denotes the ordinary time derivative, $H= \dot{a}/a$ is the
usual Hubble parameter and  the extrinsic term $B=\frac{\dot b}{b}$ is written in analogy with Hubble parameter.

\subsection{The dynamics for extrinsic curvature}
It is worth noting that eq.(\ref{6}) is not perturbed with an additional variable or parameter as commonly seen in most brane-world models. In a different approach, we use the smooth deformations concept based on Nash's embedding theorem where the perturbations can be generally performed on the spacetime itself. It means that the embedded geometry can be warped, bend or stretched. According to Nash theorem \citep{Nash} the extrinsic curvature generates the perturbations of the gravitational field along the extra
dimensions. This lends the physical interpretation that the extrinsic curvature is an independent spin-2 field in the embedded space-time.

On the other hand, in the five-dimensional bulk an additional equation for the extrinsic curvature is required
for two reasons: First, the confinement of gauge fields implies that the gravitational vector equations eq.(\ref{5}) are homogeneous and the function $b(t)$
does not have a unique solution. Therefore, they are not sufficient to determine completely the extrinsic curvature. Secondly, in the particular case of Minkowski space-time we cannot start Nash's perturbations because the extrinsic curvature of that space-time is zero.

To remedy this situation, a dynamic theory of extrinsic curvature $k_{\mu\nu}$ was proposed \citep{gde}. In his paper,  Gupta \citep{Gupta}  found  that any spin-2  field  in  Minkowski  space-time must   satisfy   an equation   that   has  the  same  formal  structure  as  Einstein's equations. This result can be obtained by   an infinite  sequence of  infinitesimal perturbations  of a linear  gravitational  equation. Accordingly, he found an  Einstein-like system of equations.

As a tentative of extension of Gupta's theorem to a curved spacetime, we found that we could not interpret the extrinsic curvature itself like a metric tensor. The choice of
the metric tensor was made as the original first fundamental form $g_{\mu\nu}$. Thus, a new tensor field $f_{\mu\nu}$ was requested to normalize $k_{\mu\nu}$ with process analogous to the ``Levi-Civita'' connection associated with $f_{\mu\nu}$. As a result, the application of Gupta's theorem led to a ``new'' geometry for spin-2 field and one can define,
\begin{equation}\label{18}
f_{\mu\nu}= \;\frac{2}{K}\;k_{\mu\nu}\;,f^{\mu\nu}=
\;\frac{2}{K}\;k^{\mu\nu}\;,\;f_{\mu\nu}f^{\mu\rho}=\;\delta^{\rho}_{\nu}\;,
\end{equation}
where $K=K(t)$ is a time-coordinate function.

Hence, an $f$-connection can be constructed and it can be given by
$$\Upsilon_{\mu\nu\sigma}=\;\frac{1}{2}\left(\partial_\mu\;f_{\sigma\nu}+\partial_\nu\;f_{\sigma\mu}
-\partial_\sigma\;f_{\mu\nu}\right)\;,\;\Upsilon^{\lambda}_{\mu\nu}= f^{\lambda\sigma}\;\Upsilon_{\mu\nu\sigma}\;.$$
It is important to note that the ``metric'' tensor $f_{\mu\nu}$ and its
inverse are used to lower and raise indices on components in the
$f$-geometry. Moreover, a ``curvature'' tensor for $f_{\mu\nu}$ as
$$\mathcal{F}_{\nu\alpha\lambda\mu}=\;\partial_{\alpha}\Upsilon_{\mu\lambda\nu}-\;\partial_{\lambda}\Upsilon_{\mu\alpha\gamma}+
\Upsilon_{\alpha\sigma\mu}\Upsilon_{\lambda\nu}^{\sigma}-\Upsilon_{\lambda\sigma\mu}\Upsilon_{\alpha\nu}^{\sigma}\;,$$
where $\partial$ denotes the ordinary derivative. Therefore, one can write the ``Ricci tensor'' and the ``scalar tensor'', respectively as
$$f^{\alpha\lambda}\mathcal{F}_{\nu\alpha\lambda\mu}=\;\mathcal{F}_{\mu\lambda\nu}^{\;\;\;\lambda}\;=\mathcal{F}_{\mu\nu}\;,
\;\mathcal{F}=\;f^{\mu\nu}\mathcal{F}_{\mu\nu}\;\;\;.$$

In addition, using the contracted Bianchi identities, one can find Einstein-Gupta's equations
\begin{equation}\label{22}
 \mathcal{F}_{\mu\nu}-\frac{1}{2}\mathcal{F} f_{\mu\nu}+\Lambda_f\;f_{\mu\nu}=\;\alpha_f\zeta_{\mu\nu}\;\;,
\end{equation}
where $\zeta_{\mu\nu}$ is the source of the field $f_{\mu\nu}$, $\Lambda_f$ and $\alpha_f$ are the cosmological
and coupling constants, respectively. It is important to stress that the $f_{\mu\nu}$ metric-type tensor is a geometric field
that only exists when the extrinsic geometry is considered. In the 4-dimensional physics such field does not exist for a riemannian
observer that only takes into account the tangent vector components. As it happens, to describe a pure gravitational spin-2 system, one can write the Einstein-Gupta's equations as
\begin{equation}\label{23}
\mathcal{F}_{\mu\nu}=\;0\;\;.
\end{equation}

\section{Statefinders analysis }
When applied to FLRW model in 5-dimensions, Gupta's equation leads to a modified Friedmann equation
\begin{equation}\label{eq:equacao de Friedman modificada}
\left(\frac{\dot{a}}{a}\right)^2+\frac{\kappa}{a^2}=\frac{8}{3}\pi
G\rho+\frac{\Lambda}{3}+\frac{b^2}{ a^4}\;\;.
\end{equation}
The contribution of the extrinsic curvature, $b(t)$, is given by
\begin{equation}\label{eq:solucao geral b(t)}
b(t)= \alpha_0\left(a\right)^{\beta_0}e^{\mp \frac{1}{2}\gamma(t)}\;\;,
\end{equation}
where all integration constants are combined in $\alpha_0$ and $\beta_0$. The term $\gamma(t)$ is denoted by
\begin{equation}\label{eq:gamma}
\gamma(t)= \sqrt{4\eta_0a^4 - 3}-
\sqrt{3}\arctan\left(\frac{\sqrt{3}}{3}\sqrt{4\eta_0a^4
- 3}\right)\;.
\end{equation}

Alternatively, the modified Friedman equation can be written as
\begin{equation}\label{eq:friedman modific por gupta}
\left(\frac{\dot{a}}{a}\right)^2+\frac{\kappa}{a^2}=\frac{4}{3}\pi
G\rho+\frac{\Lambda}{3}+\kappa_0a^{2\beta_0-4}e^{\pm\gamma(t)}
\end{equation}
where $\eta_0$ and $\kappa_0=\frac{b^2_0}{a^{2\alpha_0}_0}$ are integration constants. Interestingly, this solution was derived from geometrical approach only, with no assumption of exotic fluids.

\subsection{Deceleration parameter }
As a starting point, we use the usual form of the deceleration parameter $q(t)$ expressed in terms of the Hubble parameter conveniently written
in terms of the redshift $z$ as
\begin{equation}\label{eq:q(z)}
q(z) =\frac{1}{H(z)}\frac{dH(z)}{dz}(1 + z) - 1\;,
\end{equation}
where $H(z)$ is the Hubble parameter.

Since we are not considering $\Lambda$ as the main cause of the accelerated expansion, in the following, we neglect the cosmological constant $\Omega_{\Lambda}$ and $\Omega_{\kappa}$ curvature density parameters, and  write
\begin{equation}\label{eq:2}
H(z)=H_0\sqrt{\Omega_{\;m}(1+z)^3 +
\Omega_{\;ext}(1+z)^{4-2\beta_0}e^{\pm\gamma(z)}}\;,
\end{equation}
where $H_0$ is the current Hubble constant. It is important to note that we have two cases to study according to eq.(\ref{eq:2}) which we call for simplicity $\gamma^{+}$ and $\gamma^{-}$ solutions. The matter density parameter is denoted by $\Omega_m$. The term $\Omega_{\;ext}$ stands for the density parameter associated with the extrinsic curvature. Thus, we express (\ref{eq:gamma}) in terms of the redshift
\begin{equation}\label{eq:gammaz}
\gamma(t)= A(z)- \sqrt{3}\arctan\left(\frac{\sqrt{3}}{3}A(z)\right)\;.
\end{equation}
where $A(z)=\sqrt{\frac{4\eta_0}{(1+z)^4} - 3}$.
Notice that $\gamma(z)$ must be real so that we must have the condition
\begin{equation}\label{eq:4}
\eta_0\geq 0.75(1+z)^4\;.
\end{equation}

As a natural start point, based on the fact that the equal sign in the eq.(\ref{eq:4}) holds for $\pm \gamma(z) = 0$, this corresponds to the phenomenological solution found in \citep{GDEI} that mimics the X-CDM model with a correspondence
\begin{equation}\label{eq:exotic}
4-2\beta_0 = 3(1+ w)\;,
\end{equation}
where the parameter $w$ holds for the exotic fluid parameter for an X-fluid \citep{turner}. For instance, motivated by the latest Planck observations, for $\Lambda CDM $ (with $w=-1$) and phantom models ($w<-1$) we have $\beta_0 =2$ and $\beta_0 =3$, respectively. It is worth noting that this correspondence is not a necessary condition (but an easier way) to constrain the parameters. This first prior constraint can avoid a larger error propagation from observational data (in the case of using $\chi$-square fitting as shown in the study of high-order statefinders \citep{Lazkoz}). We are interested in solutions $\pm \gamma \neq 0$ in the low-redshift $[0,2.3]$ where we have more reliable data.

In order to test the model, we need to constraint the two parameters $(\eta_0,\beta_0)$. In the course of this study, we notice that $\beta_0$ affects the value of current deceleration parameter $q_0$ and $\eta_0$ rules mainly on the width of the transition phase $z_t$. For the $\beta_0$ parameter, using eq.(\ref{eq:exotic}) we have found the constraint
\begin{equation}\label{eq:constraint01}
    2 \leq \beta_0 \leq 3\;.
\end{equation}

To constrain the values of $\eta_0$, we impose that it must be restricted from its current value (i.e, for the redshift $z=0$) to the asymptotic value of the deceleration parameter (i.e, for the redshift $z=-1$). In these terms we have
\begin{equation}\label{eq:constraint02}
    0 \leq \eta_0 \leq 0.75\;.
\end{equation}
This result does not contradict eq.(\ref{eq:4}).

Concerning the cosmological density parameters and using the normalization $H\rfloor_{z=0} = H_0$, we fix the matter contribution to be $\Omega_m = 0.3$. For the present epoch,
$$\Omega_{total}\rfloor_{z=0} = \Omega_{ext}\exp(\gamma(0)) + \Omega_m = 1\;.$$
The density parameter associated with the extrinsic curvature can be given by
$$\Omega_{ext}= \frac{1- \Omega_m}{\exp(\gamma(0))}.$$
Moreover, using the definition in eq.(\ref{eq:q(z)}), we obtain the deceleration parameter
\begin{equation}\label{eq:q(z)2}
q(z)= \frac{3}{2}\left[\frac{\Omega_{\;m}(1+z)^3 + \gamma^{*}
\Omega_{\;ext}(1+z)^{4-2\beta_0}e^{\mp \gamma(z)}}
{\Omega_{\;m}(1+z)^3+\Omega_{\;ext}(1+z)^{4-2\beta_0}e^{\mp\gamma(z)}}\right]-1\;\;.
\end{equation}
where $\gamma^{*}=\frac{1}{3}\left[4-2\beta_0 \pm 2 \sqrt{\frac{4\eta_0}{(1+z)^4} - 3}\;\right]$. In addition, using eq.(\ref{eq:q(z)2}) with the constrained values of $\beta_0$ in eq.(\ref{eq:constraint01}) for both solutions $\{\gamma^{(+)}, \gamma^{(-)}\}$, we find that the values of $\eta_0\geq 0.5$ lead to an incompatible pattern as expected for deceleration parameter. Thus, the constraints on $\eta_0$ can be tighter and written as
\begin{equation}\label{eq:constraint03}
    0 \leq \eta_0 \leq 0.5\;.
\end{equation}

In table (\ref{tb:table1}), based on eq.(\ref{eq:constraint01}) and eq.(\ref{eq:constraint03}), we study six models for the cases $\{\gamma^{(+)}, \gamma^{(-)}\}$. Hereafter, in order to facilitate the visualization, all models of both $\gamma$-solutions are represented by the same type of specific curve. The $\Lambda$CDM model is represented by a solid line. Accordingly, Models I, II, III, VI, V and VI are presented by lines with short-dot, dash, dash-dot-dot with an up-triangle, dash-dot with a circle, dot and dash with a star, respectively. It is important to point out that what was intended in defining each model, with a certain value of $(\eta_0,\beta_0)$, was varying its possible values with its respective ranges as shown in eqs.(\ref{eq:constraint03}) and (\ref{eq:constraint01}). For instance, the models V and VI are defined with $\eta_0=0.1$ and $\beta_0=2.5$ and $\beta_0=3$, respectively. Varying $\eta_0$ from $0.1$ to $0.5$, we do not verified a considerable difference in the results. In this respect, with the proposed models in course, one can expect to obtain a general view of the applicability of this theoretical scheme.

\begin{table}
\centering
   \caption{Six models studied for selected values of $(\eta_0, \beta_0)$ and the predicted current deceleration parameter $q_0$ for both $\gamma^{(+)},\gamma^{(-)}$ solutions.}
  \begin{tabular}{@{}llrrrlllll@{}}
  \hline
  $\gamma$-solutions      &            &           &$\gamma^{(+)}$  &   $\gamma^{(-)}$ \\
  Parameters              & $\eta_0$   & $\beta_0$ & $q_0$          &    $q_0$          \\
  Model                   &            & \\
  \hline
  I           & 0.001    & 2      & -0.549      &   -0.556 \\
  II          & 0.1      & 2      & -0.460      &   -0.640 \\
  III         & 0.25     & 2      & -0.341      &   -0.759 \\
  IV          & 0.5      & 2      & -0.362      &   -0.738  \\
  V           & 0.1      & 2.5    & -0.810      &   -0.996 \\
  VI          & 0.1      & 3      & -1.160      &   -1.340 \\
  \hline
\end{tabular}\label{tb:table1}
\end{table}

\begin{figure*}
\centering
\mbox{\subfigure{\includegraphics[width=3.2in, height=4in]{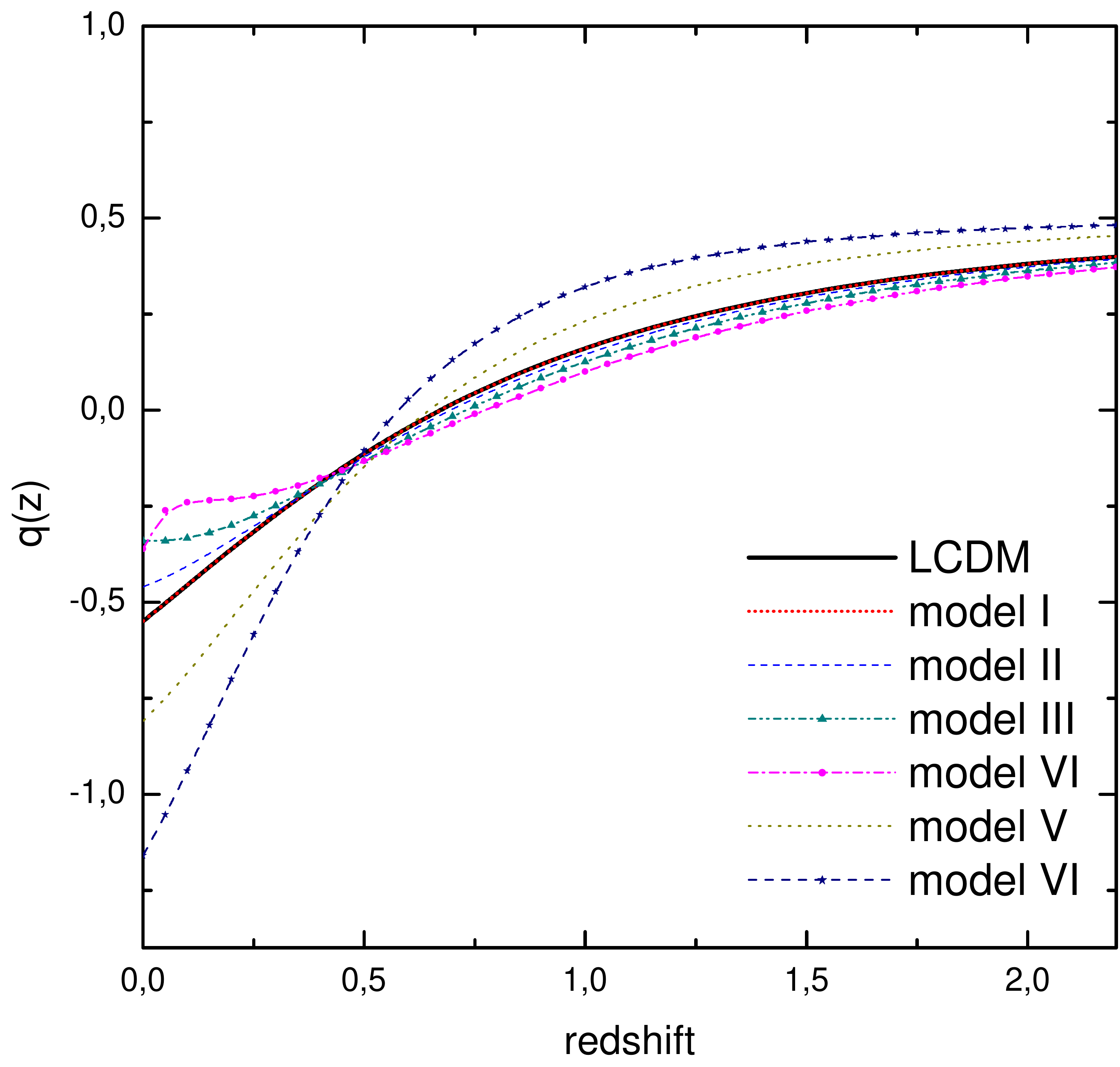}}\quad
\subfigure{\includegraphics[width=3.2in, height=4in]{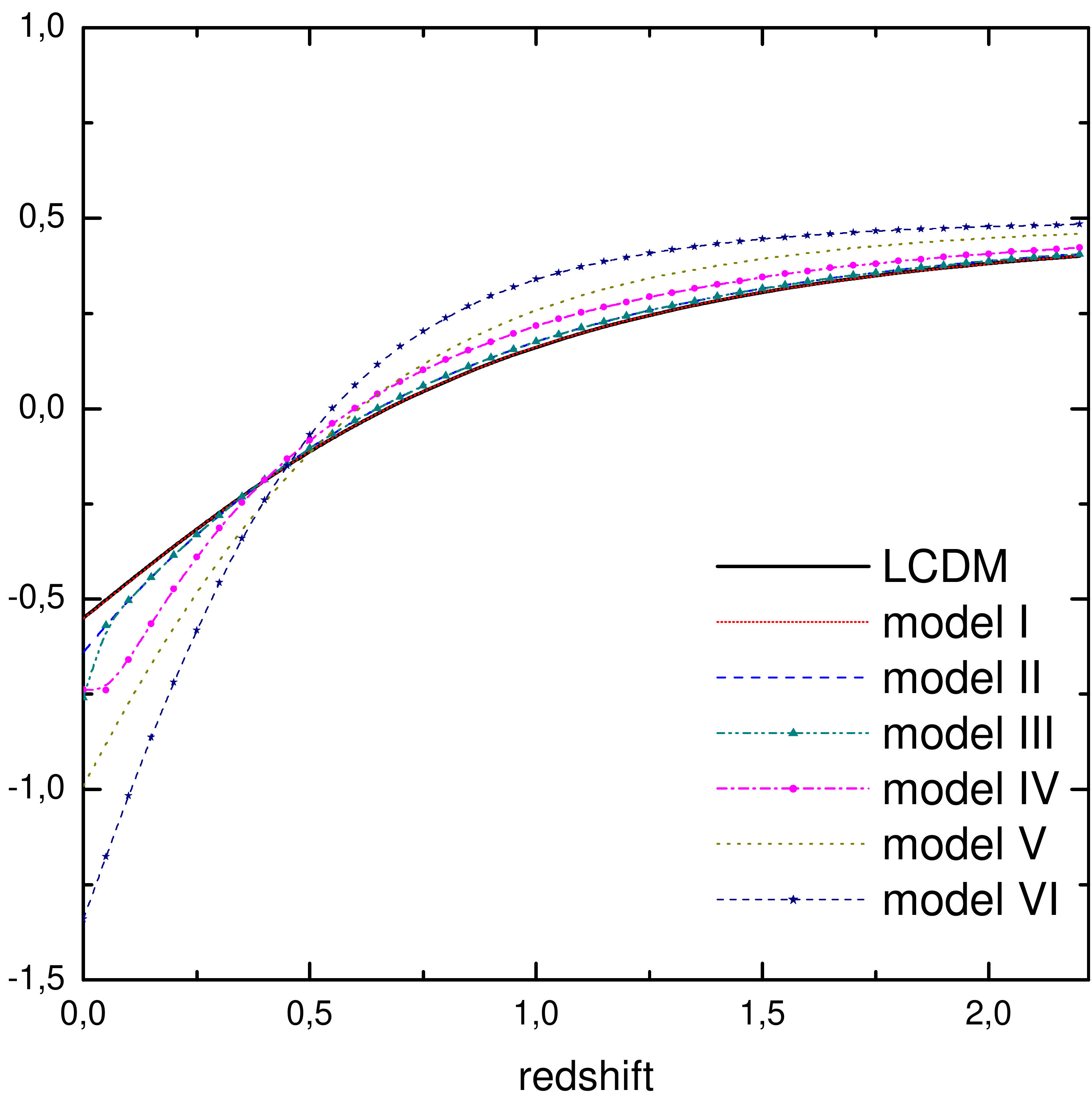} }}
\caption{Deceleration parameter as a function of redshift for solutions of $\gamma^{(+)}$ (left figure) and $\gamma^{(-)}$ (right figure). From each model, the resulting current deceleration parameter $q_0$ is also indicated.} \label{fig:q}
\end{figure*}

The resulting deceleration parameter for $\gamma^{(+)}$ is shown in the left figure in fig.(\ref{fig:q}).

The model I is in concordance with $\Lambda$CDM (solid line). It is worth noting that the models from II to V  are also in concordance with BAO/CMB + SNIa \citep{Giostri}. The model VI presents a phantom-like pattern. For these models, the transition redshift $z_t$ lies at $0.57 \leq z_t \leq 0.74$.

On the other hand, the $\gamma^{-}$ solution predicts models with more intense acceleration as compared with the previous case as it can be shown in the right figure in fig.(\ref{fig:q}). Again, the model I is in completely concordance (overlapped) with $\Lambda$CDM and no significant differences can be observed. It is worth noting that the models I to V  are in concordance with BAO/CBM + SNIa \citep{Giostri}. The model VI presents a phantom-like pattern. For these models, the transition redshift $z_t$ lies at $0.55 \leq z_t \leq 0.68$.

According with BAO/CMB + SNIa constraints, with MLCS2K2 light-curve fitter it gives $z_t= 0.56^{+0.13}_{-0.10}$ and $q_0= -0.31^{+0.11}_{-0.11}$ with $68\%$ C.L favors DGP-like models. With  SALT2 fitter, $z_t= 0.64^{+0.13}_{-0.07}$ and $q_0= -0.53^{+0.17}_{-0.13}$ with $68\%$ C.L that favors $\Lambda$CDM models. For both solutions $\{\gamma^{(+)},\gamma^{(-)}\}$, the transition $z_t$ mimics DGP-like models as well as $\Lambda$CDM models. As table (1) indicates for $\gamma^{(+)}$ solution, models I and II mimic $\Lambda$CDM-like models. The model III and IV mimic DGP-like models. Finally, the model V and VI indicate a quintessence-like and phantom-like behavior, respectively.

In addition, for $\gamma^{(-)}$ solution, models I and II also mimic $\Lambda$CDM-like models. Differently to the previous $\gamma^{(+)}$ case, for the $\gamma^{(-)}$ solution for the models III, IV and V, they mimic a quintessence-like models. Finally, the model VI also indicates a phantom-like behavior just like in the case of $\gamma^{(+)}$.

It is worth noting that in both cases $\{\gamma^{(+)},\gamma^{(-)}\}$, the models I and II mimic $\Lambda$CDM-like models according with BAO/CMB + SNIa constraints. This was expected with $\beta_0=2$ and a small $\eta_0$ close to zero. A similar expectation have occurred to obtain quintessence-like and/or phantom-like behaviors when we set $\beta_0$ close or equal to 3 which we have confirmed with the models V and VI (for both $\gamma$-solutions) with a more negative current value for the deceleration parameter. On the other hand, in order to select models even more constrained, we need to analyze the jerk parameter as shown in the next section.

\subsection{Jerk parameter }
The jerk parameter $j$ is defined as the third time-order derivative of the scale factor \emph{a} and is given by
\begin{equation}\label{eq:jerk}
    j= \frac{\dot{\ddot{a}}}{aH^3}\;.
\end{equation}

If one takes a Taylor expansion of the scale factor around its current value $a_0$,
\begin{eqnarray*}
&&\frac{a(t)}{a_0}= 1 + H_0 (t-t_0) - \frac{1}{2}q_0 H_0^2 (t-t_0)^2 +\\
&&\hspace{2.5cm} \frac{1}{6}j_0 H_0^3 (t-t_0)^3 +\mathcal{O}[(t-t_0)^4]+... ,
\end{eqnarray*}
and eq.(\ref{eq:jerk}) can be rewritten as \citep{Poplawski}
\begin{equation}\label{eq:jerk2}
    j = q + 2 q^2 - \frac{1}{H} \frac{dq}{dz}\;,
\end{equation}
where we denote $j=j(z)$ and $q=q(z)$.

\begin{figure*}
\centering
\mbox{\subfigure{\includegraphics[width=3.2in, height=4in]{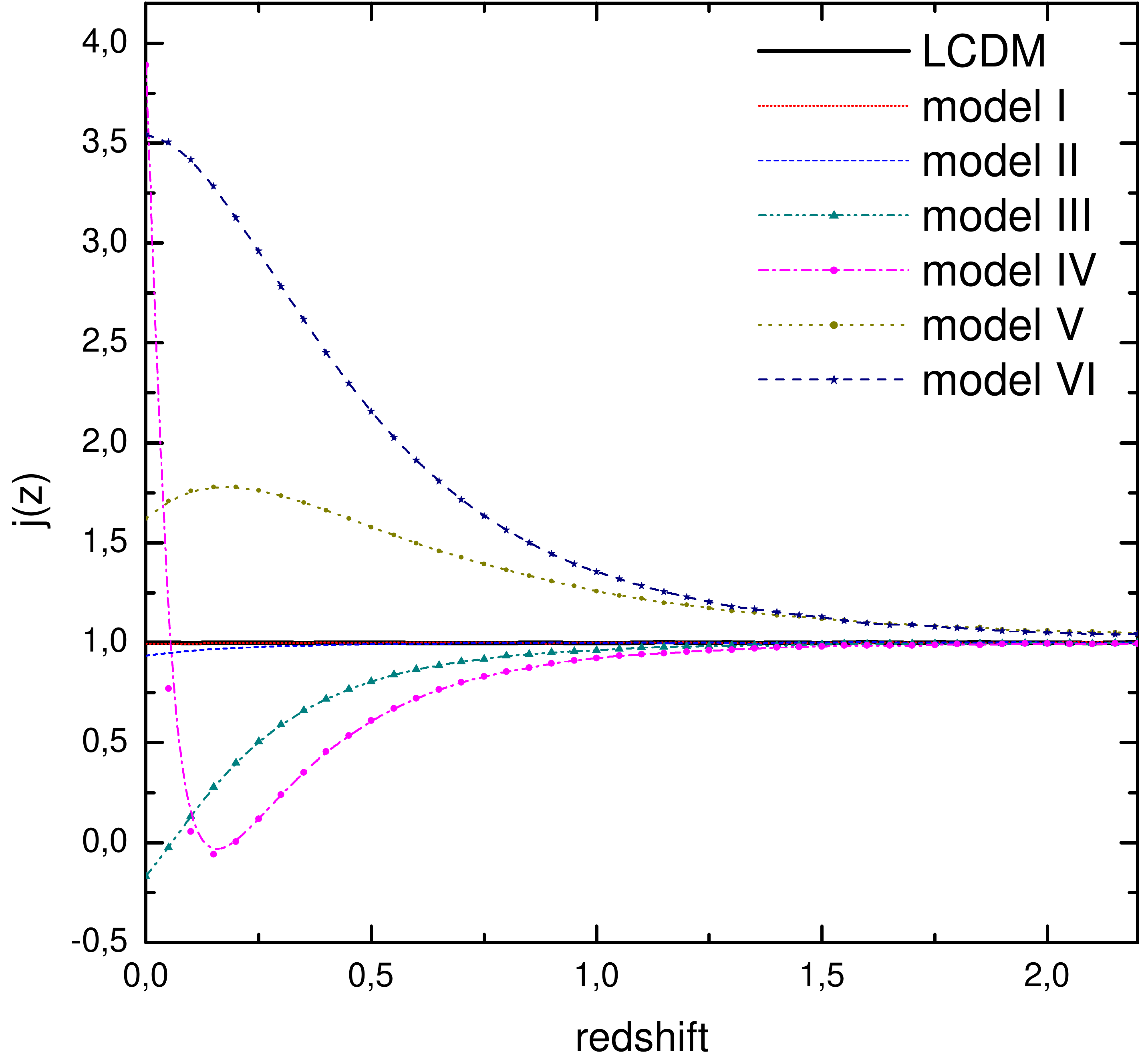}}\quad
\subfigure{\includegraphics[width=3.1in, height=4.1in]{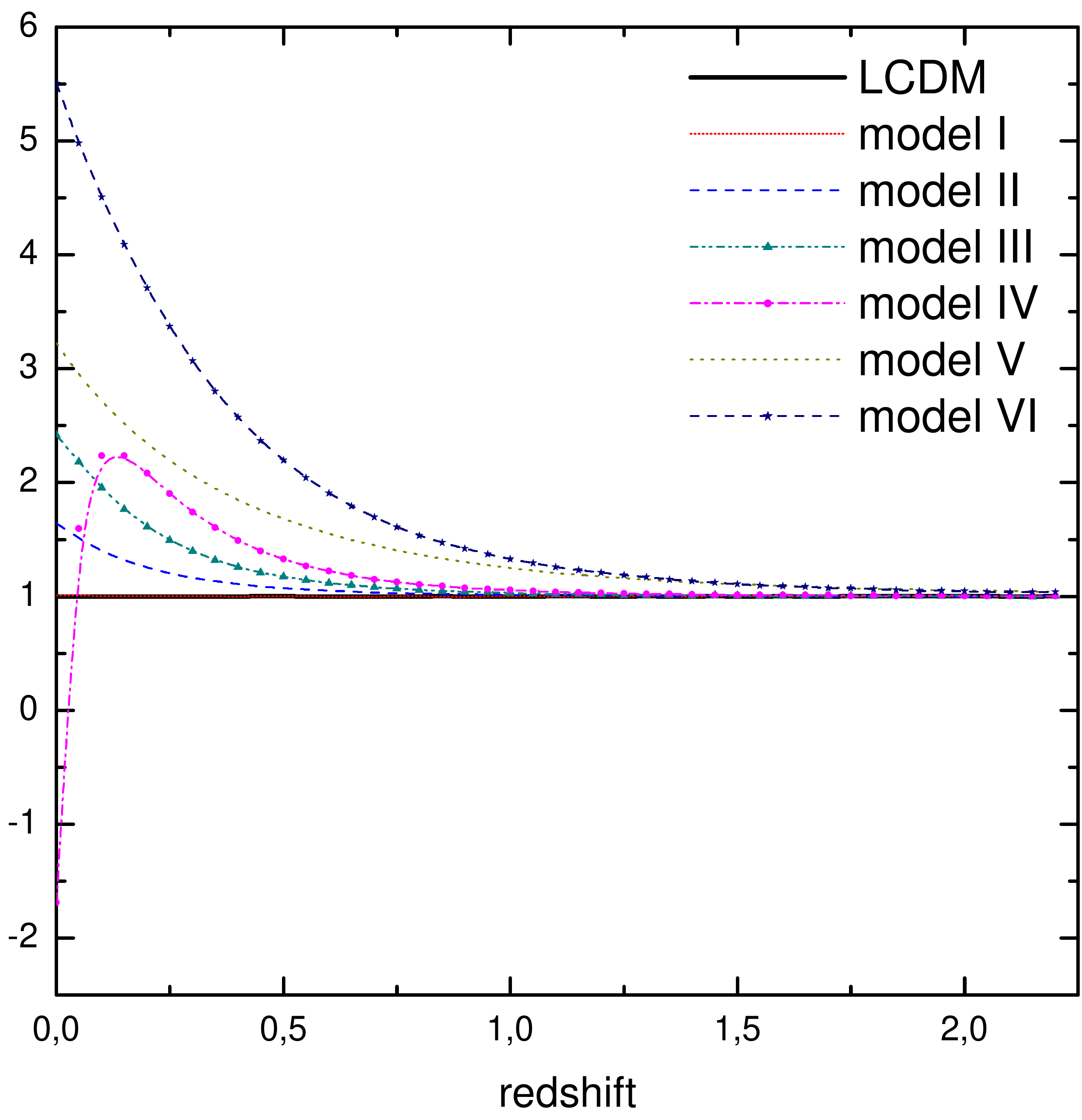} }}
\caption{Jerk parameter as a function of redshift for solutions of $\gamma^{(+)}$ (left figure) and $\gamma^{(-)}$ (right figure).} \label{fig:jerk}
\end{figure*}

Since the hubble parameter $H(z)$ depends on the differential age as a function of the redshift such as
$$H(z) =  - \frac{1}{1+z} \frac{dz}{dt}\;,$$
one can write the jerk parameter in terms of redshift as
\begin{equation}\label{eq:jerk3}
    j = q + 2 q^2 + (1+z) \frac{dq}{dz}\;.
\end{equation}

Another feature of the jerk parameter is that it can be used to select gravitational models. Using  eq.(\ref{eq:jerk3}), we obtain fig.(\ref{fig:jerk}) for the solutions  $\gamma^{(+)}$ and $\gamma^{(-)}$, respectively. The pair $(\eta_0, \beta_0)$ for each $\gamma$ solutions are the same as used in table (\ref{tb:table1}).

We compare now with the SNLS SNIa dataset that gives $j = 1.32^{+1.37}_{-1.21}$ \citep{astier}
x-ray galaxy clusters $j = 0.51^{+2.55}_{-2.00}$\citep{rapetti} and the gold sample (SNIa)\citep{riess} that gives $j = 2.75^{+1.22}_{-1.10}$.
For $\gamma^{(+)}$ solution in the left figure in fig.(\ref{fig:q}), the models I, II and V are compatible with all the previous datasets. With a possible negative jerk, model 3 are compatible with the x-ray galaxy cluster only. The model VI are compatible with the gold sample (SNIa) only. For $\gamma^{(-)}$ solution in the right figure in fig.(\ref{fig:q}), the models I, II and III are compatible with all the previous datasets. The model V is compatible with both x-ray galaxy clusters and the gold sample (SNIa). The model VI is incompatible with observations and the phantom-like behavior is not allowed, at least in $\gamma^{(-)}$ solution.

The model IV fails in both $\gamma$-solutions and can be neglected. A common feature of all models is that for redshift $z<2$ all the curves tend to the $\Lambda$CDM-like pattern.

\subsection{Hubble parameter }
\begin{figure*}
\centering
\mbox{\subfigure{\includegraphics[width=3.3in, height=4in]{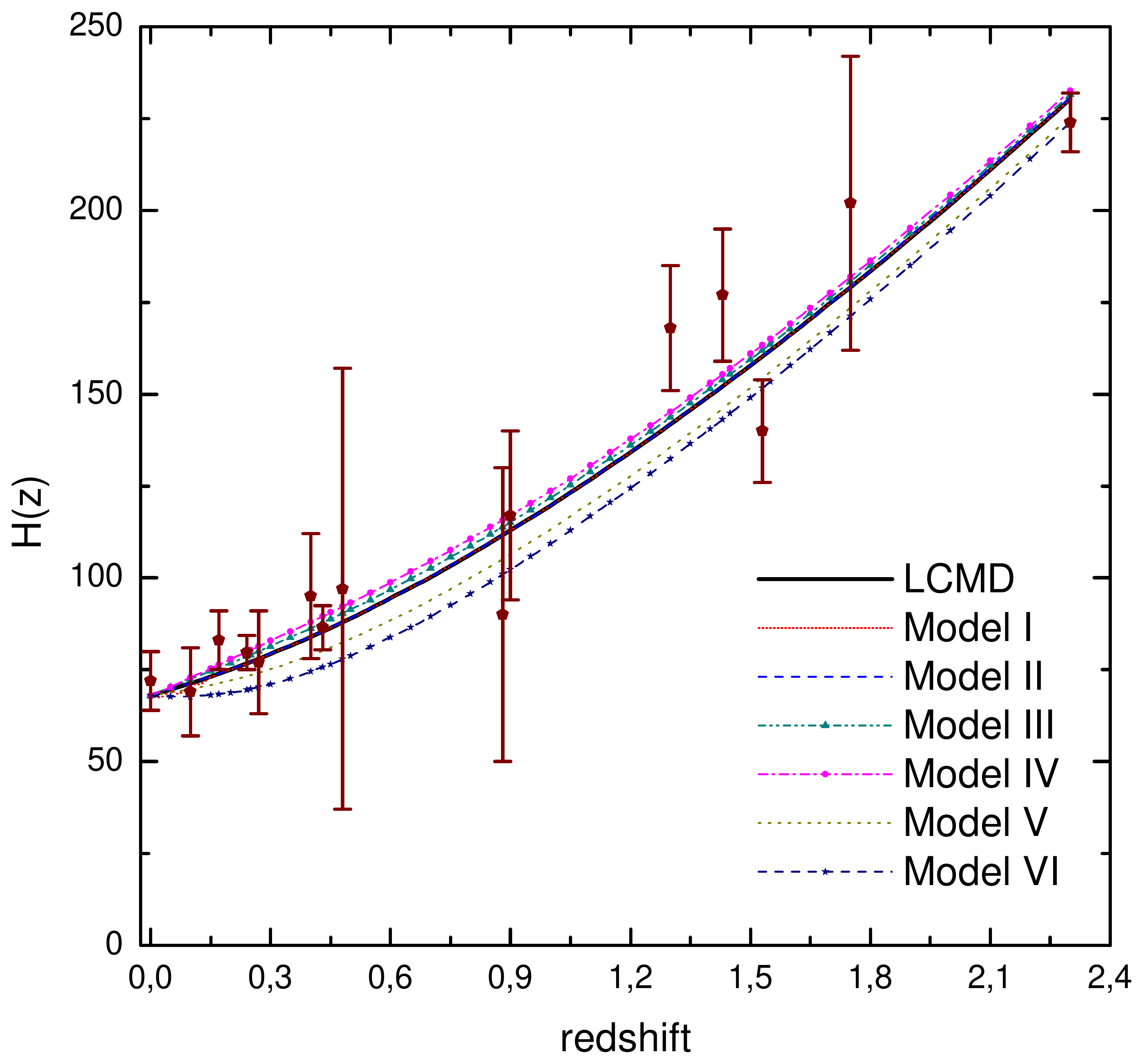}}\quad
\subfigure{\includegraphics[width=3.2in, height=4in]{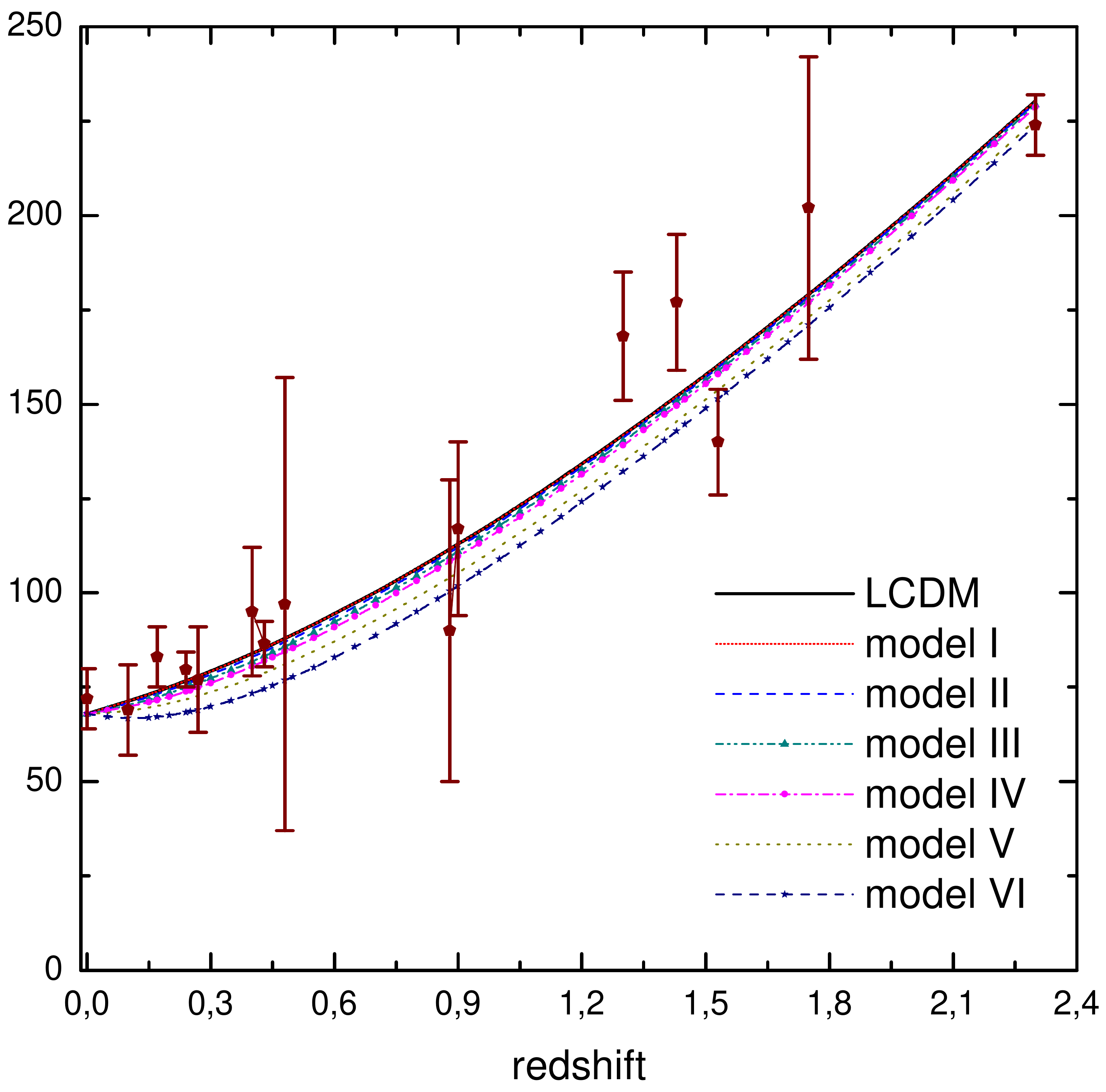} }}
\caption{Deceleration parameter as a function of redshift for solutions of $\gamma^{(+)}$ (left figure) and $\gamma^{(-)}$ (right figure). Error bar points were extracted from \citep{cao}  supplemented with additional from \citep{busca,zhai} at $H(z = 2.3)$.}\label{fig:hub}
\end{figure*}

We also study the evolution of the hubble parameter $H(z)$ in the redshift range $0 < z< 1.8$ with original data observational points extracted from \citep{cao} based on observations of red-enveloped galaxies \citep{stern} and BAO peaks \citep{gaztanaga}. It was also supplemented with the observational Hubble parameter data (OHD) and BAO in Ly$\alpha$ \citep{busca,zhai} implying that $H(z = 2.3) = (224 \pm 8) km.s^{-1}.Mpc^{-1}$. In this respect, we obtain the following curves in the fig.(\ref{fig:hub}).

For both cases $\gamma^{(+)}$ and $\gamma^{(-)}$, we obtain a similar pattern very close to $\Lambda$CDM-like behavior for the first four models of both $\gamma$-solutions. In  $\gamma^{(-)}$ solution, the model III and IV indicate a lower values for Hubble parameter as compared with the same models in $\gamma^{(+)}$ solution. This characteristic can be interesting to further studies on primordial nucleosynthesis. The models V and VI present the lowest values for Hubble parameter as compared with all models studied but still with agreement with observations \citep{busca}. The overall conclusion is due to the results of the curves of jerk parameters, the model IV must be neglected in both $\gamma$-solutions which means that the $\eta_0$ parameter can be constrained to the range $0 \leq \eta_0 < 0.5$.In the same sense, the model VI in $\gamma^{(-)}$ solution can be neglected. Moreover, the other models present a good concordance with observations and can be explored as a package for cosmological purposes and more constrained with  improved techniques of near-future observations.

\section{Final remarks}
We  have  applied  the   concept  of  smooth   deformations of  riemannian  manifolds  to  space-times  and in  particular  to  the  FLRW  universe,  showing  that it is  an  efficient  mechanism  to   explain  the  accelerated  expansion of  the  universe. In such geometric process, some of the topological informations of space-times does not appear in riemannian geometry. For instance, the  conserved  quantity  $Q_{\mu\nu}$  may  be   interpreted  as  a  component of some  mechanical  energy  responsible   for  the
observed  acceleration of  the  universe.  However,  from the  point of  view of  geometry,   it  may  be  also  interpreted   as a  necessary   observational   quantity,   reintroducing  some  topological  qualities to a gravitational theory.  This  latter  interpretation  gives  full    support  to the  Gauss  and  Riemann views  that   geometry   is  determined  essentially  by  the  observations,  regardless  how  small and  near  or  how  large  and  distant as  they  may  be.

The acceleration of the universe was described as a consequence of the extrinsic curvature of the space-time
embedded in a bulk space, defined by the Einstein-Hilbert action. It seems natural that this result provides the required geometrical structure to describe a dynamically changing universe. The four-dimensionality of the embedded space-times was determined by the dualities of the gauge fields,
which corresponds to the equivalent concept of confinement gauge fields and ordinary matter in the brane-world
program. However, this confinement implies that the extrinsic curvature cannot be completely determined,
simply because Codazzi's equations becomes homogeneous. Since the extrinsic curvature assumes a fundamental
role in Nash's theorem, an additional equation was required. We have noted that the extrinsic curvature is an
independent rank-2 symmetric tensor which corresponds to a spin-2 field defined on the embedded space-time.
However, as it was demonstrated by Gupta, any spin-2 field must satisfy an Einstein-like equation.

After the due adaption to an embedded space-time, in a previous work \citep{gde} we have constructed Einstein-Gupta's equations for the extrinsic curvature
of the FLWR geometry. Here we have extended the idea to the study of the behavior of its solution at low redshift. This was compared with a phenomenological model (XCDM), obtaining very similar patterns but using a geometrical approach. However, two models in $\gamma^{(\pm)}$-solutions were neglected constrained in the analysis of the jerk parameter revealing a incompatibility with the observational data.

It is important to stress that all different datasets presented were not used to obtain the results presented in this paper. On the contrary, they were used to constrain the models studied only. As future perspectives, the models presented must be studied with the analysis on high redshift and primordial nucleosynthesis. Also, the snap and lerk parameters must be contemplated and should be investigated further.

\label{lastpage}

\end{document}